\documentclass[prb,twocolumn,superscriptaddress,showpacs,showkeys]{revtex4}
\usepackage[latin1]{inputenc}
\usepackage{graphicx}
\usepackage{amsmath,amssymb}
\usepackage{dcolumn}
\usepackage{bm}
\usepackage{epsfig}
\usepackage{longtable}
\usepackage{subfigure}
\usepackage{multirow}

\newcommand{\kpe}{\mathbf{k}\!\cdot\!\mathbf{p}\,}
\newcommand{\Kpe}{\mathbf{K}\!\cdot\!\mathbf{p}\,}
\newcommand{\bfr}{ {\bf r}} 
\newcommand{\bfrp}{ {\bf r'}} 
\newcommand{\bfR}{ {\bf R}} 
\newcommand{\bfq}{ {\bf q}} 
\newcommand{\bfk}{ {\bf k}} 

\newcommand{\bracket}[3]{\ensuremath{\langle #1 | #2 | #3 \rangle}}
\newcommand{\GnWn}{\ensuremath{G_0W_0}\,}
\newcommand{\tn}[1]{\textnormal{#1}}
\newcommand{\f}[1]{\footnotemark[#1]}

\newcommand{\mcs}[3]{\multicolumn{#1}{#2}{#3}}
\newcommand{\mcc}[1]{\multicolumn{1}{c}{#1}}
\newcommand{\gw}{$\tn{OEPx(cLDA)}$+$\GnWn$}
\allowdisplaybreaks
\begin{document}

\title{{\rm\small\hfill (accepted at Phys.\ Rev.\ B)}\\
        Consistent set of band parameters for the group-III nitrides AlN, GaN, and InN}
\date{\today}

\author{Patrick Rinke}
        \email{rinke@fhi-berlin.mpg.de}
	\affiliation{Fritz-Haber-Institut der Max-Planck-Gesellschaft, Faradayweg~4-6,D-14195 Berlin, Germany}
\author{M.~Winkelnkemper}
	\affiliation{Fritz-Haber-Institut der Max-Planck-Gesellschaft, Faradayweg~4-6,D-14195 Berlin, Germany}
	\affiliation{Institut für Festk\"{o}rperphysik, Technische Universit\"{a}t Berlin, Hardenbergstra{\ss}e~36, D-10623 Berlin, Germany}
\author{A.~Qteish}        
        \affiliation{Department of Physics, Yarmouk University, 21163-Irbid, Jordan}
\author{D.~Bimberg} 
	\affiliation{Institut für Festk\"{o}rperphysik, Technische Universit\"{a}t Berlin, Hardenbergstra{\ss}e~36, D-10623 Berlin, Germany}
\author{J.~Neugebauer}
        \affiliation{Max-Planck-Institut fur Eisenforschung, Department of Computational Materials Design, D-40237 D\"{u}sseldorf, Germany}
\author{M.~Scheffler}
	\affiliation{Fritz-Haber-Institut der Max-Planck-Gesellschaft, Faradayweg~4-6,D-14195 Berlin, Germany}

\begin{abstract}
We have derived consistent sets of band parameters (band gaps, crystal field-splittings, 
band gap deformation potentials, effective masses, Luttinger and $E_\tn{P}$ parameters) 
for AlN, GaN, and InN in the zinc-blende and wurtzite phases employing many-body perturbation 
theory in the \GnWn approximation.  
The \GnWn method has been combined with density-functional theory (DFT) calculations in the 
exact-exchange optimized effective potential approach (OEPx) to overcome the limitations of 
local-density or gradient-corrected DFT functionals (LDA and GGA).  
The band structures in the vicinity of the $\Gamma$-point have been used to directly 
parameterize a 4\,x\,4 $\kpe$ Hamiltonian to capture non-parabolicities in the conduction 
bands and the more complex valence-band structure of the wurtzite phases.  
We demonstrate that the band parameters derived in this fashion are in very good agreement 
with the available experimental data and provide reliable predictions for all parameters 
which have not been determined experimentally so far.

\end{abstract}




\maketitle

\section{\label{sec:introduction}Introduction}
        The group III-nitrides AlN, GaN, and InN and their alloys have become an important and
versatile class of semiconductor materials, in particular for use in optoelectronic devices 
and high-power microwave transistors.
Current applications in solid state lighting [light emitting diodes (LEDs) and laser diodes (LDs)] range from the visible spectrum 
\cite{nakamura1991,nakamura1998,nakamura1996_1,nakamura1996_2} to the deep ultra-violet (UV)\cite{taniyasu2006,adivarahan2004}, while future applications as, e.g., chemical
sensors \cite{Stutzmann/etal:2002,Schalwig/etal:2002,Lu/etal:2004} or in quantum cryptography \cite{kako2006} are being explored.

For future progress in these research fields 
reliable material parameters beyond the fundamental band gap, like effective electron masses and valence-band (Luttinger or Luttinger-like) parameters, are needed to aid interpretation of experimental observations and to simulate (hetero-)structures, like, e.g., optoelectronic devices. 
Material parameters can be derived from first-principles electronic-structure methods for bulk phases, but the size and complexity of structures required for device simulations
currently exceeds the capabilities of first-principles electronic-structure tools by far.
To bridge this gap first-principle calculations can be used to parameterize simplified methods, like the $\kpe$ method,\cite{kane1982,andreev2000,fonoberov2003,winkelnkemper2006} the empirical tight-binding (ETB) method,\cite{slater1954,saito2002, ranjan2003, schulz2006} or the empirical pseudo-potential method (EPM) \cite{wang1993}, which are applicable to large-scale heterostructures at reasonable computational expense.

In this Article we use many-body perturbation theory in the \GnWn approximation,\cite{Hedin:1965}---currently the method of choice for the description 
of quasiparticle band structures in solids \cite{Aulbur/Jonsson/Wilkins:2000,Onida/Reining/Rubio:2002,Rinke/etal:2005}---in combination with the $\kpe$ approach,
\cite{kane1982,andreev2000,fonoberov2003,winkelnkemper2006} to derive a consistent set of material parameters for the group-III nitride system.
 The $\kpe$-Hamiltonian is parameterized to reproduce the $\GnWn$ band structure in the vicinity of the $\Gamma$-point. Since the parameters of the $\kpe$ method are closely related or, in some cases, even identical to basic band parameters, many key band parameters can be directly obtained using this approach. 

The $\kpe$-model Hamiltonian is typically parameterized for bulk structures and is then 
applicable to heterostructures with finite size (e.g., micro- and nanostructures) within the envelope-function 
scheme.\cite{bastard1986}  Ideally, the parameters are determined entirely from consistent 
experimental input.  For the group-III-nitrides, however, many of the key band parameters 
have not been conclusively determined until now, despite the extensive research effort in this 
field.\cite{Walukiewicz/review:2006,Vurgaftman/Meyer:2003}  In a comprehensive review 
Vurgaftman and Meyer summarized the field of III-V semiconductors in 2001 and recommended 
up-to-date band parameters for all common compounds and their alloys including the nitrides.
\cite{vurgaftman2001}
Only two years later they realized that \emph{it is striking how many of the nitride 
properties have already been superseded, not only quantitatively but 
qualitatively}.\cite{Vurgaftman/Meyer:2003} They proceeded to 
\emph{remedy that obsolescence, by providing a completely revised and updated description 
of the band parameters for nitride-containing semiconductors} in 
2003.\cite{Vurgaftman/Meyer:2003} While this update includes evidence supporting a 
revision of the band gap of InN from its former value of 1.9\,eV to a significantly lower 
value around 0.7\,eV,
\cite{Davydov/etal:2002,Wu/etal:2002_1,Nanishi/etal:2003, Sher/etal:1999,
      Bechstedt/Furthmueller:2002} they had to concede that in many cases experimental 
information on certain parameters was simply not available.\cite{Vurgaftman/Meyer:2003}
This was mostly due to growth-related difficulties in producing high quality samples for unambiguous characterization. In the meantime the quality of, e.g., wurtzite InN samples has greatly improved \cite{Walukiewicz/review:2006} and even the growth of the zinc-blende phase has advanced.\cite{Lozano/etal:2007} Nevertheless, many of the basic material properties of the group-III nitrides are still undetermined or, at least, controversial. 

On the theoretical side, certain limitations of density-functional theory (DFT) in the 
local-density or generalized gradient approximation (LDA and GGA, respectively)---currently 
the most wide-spread {\it ab-initio} electronic-structure method for 
poly-atomic systems---have hindered an unambiguous completion of the missing data. 
To overcome these deficiencies we use \GnWn calculations based on
DFT calculations in the exact-exchange optimized effective potential approach (OEPx) to
determine the basic band parameters. We have previously shown that the OEPx+\GnWn approach
provides an accurate description of the quasiparticle band structure for GaN, InN and II-VI
compounds \cite{Rinke/etal:2005,Rinke/etal:2006,Rinke/psi-k:2007,Rinke/pssb:2008}.
The quasiparticle band structure in the vicinity of the $\Gamma$-point is then
used to parameterize a $4\times 4$ $\kpe$ Hamiltonian to determine band-dispersion parameters,
like effective masses, Luttinger parameters, etc. 
This allows us to take the non-parabolicity of the conduction band, which is particularly
pronounced in InN \cite{Wu/etal:2002_1,Wu/etal:2002}, and the more complex valence band 
structure  of the wurtzite phases into account properly.

This paper is organized as follows: In Section \ref{sec:qpe} we briefly introduce the \GnWn 
approach and its application to the group-III nitrides, followed by a discussion of certain 
key parameters of the quasiparticle band structure, such as the fundamental band gaps 
(and their dependence on the unit-cell volume) and the crystal-field splitting energies 
(Section~\ref{sec:qpe_bs}).  In Section~\ref{sec:kp_bs} we present our recommendations for 
the band dispersion parameters ($\kpe$ parameters) of the wurtzite and zinc-blende phases of 
AlN, GaN and InN. A detailed discussion of the parameter sets is given in 
Section~\ref{sec:kp_dis} together with a comparison to experimental values and parameter 
sets obtained by other theoretical approaches.
Our conclusions are given in Sec.~\ref{sec:conc}.

\section{\label{sec:qpe}Quasiparticle energy calculations}

  \subsection{\label{sec:OEPx_GW} \textit{GW} based on
              exact-exchange DFT}

\begin{figure}[t]
   \epsfig{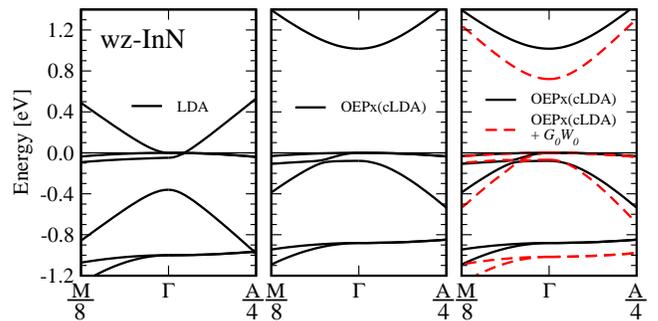}
   \caption{\label{fig:InN_BS}
	    LDA Kohn-Sham calculations incorrectly predict wurtzite InN  
	    to be a metal with the wrong band ordering at the $\Gamma$ point. 
            In OEPx(cLDA) the band gap
	    opens and InN correctly becomes a semiconductor, thus providing
	    a more suitable starting point for subsequent quasiparticle energy
	    calculations in the $G_0W_0$ approximation. }
\end{figure}

The root of the deficiencies in LDA and GGA for describing spectroscopic properties like the
quasiparticle band structure can be found in a combination of different factors. 
LDA and GGA are approximate (jellium-based) exchange-correlation functionals, 
which suffer from incomplete cancellation of artificial self-interaction and 
lack the discontinuity of the exchange-correlation potential with respect to the
number of electrons. As a consequence the Kohn-Sham (KS) single-particle eigenvalues
cannot be rigorously interpreted as the quasiparticle band structure as
measured by direct and inverse photoemission. This becomes most apparent for the band
gap, which is severely underestimated by the Kohn-Sham eigenvalue difference in LDA and GGA. 
For InN this even results in an overlap between the conduction and the valence bands and thus an
effectively metallic state, as displayed in Fig.~\ref{fig:InN_BS}. It goes without mentioning that
a $\kpe$ parameterization derived from this LDA band structure would not appropriately reflect the
properties of bulk InN.

Many-body perturbation theory in the $GW$ approach \cite{Hedin:1965}
presents a quasiparticle theory that overcomes the
deficiencies of LDA and GGA and provides a suitable description of the quasiparticle band
structure of weakly correlated solids, like AlN, GaN and InN.
\cite{Aulbur/Jonsson/Wilkins:2000,Onida/Reining/Rubio:2002,Rinke/etal:2005}
Most commonly, the Green's function $G_0$ and the screened
potential $W_0$ required in the $GW$ approach (henceforth denoted $G_0W_0$) 
are calculated from a set of DFT Kohn-Sham single particle energies and wave functions.
The DFT ground state calculation is typically carried out in the LDA or GGA and
the quasiparticle corrections to the Kohn-Sham eigenvalues are calculated in first
order perturbation theory (LDA/GGA + $\GnWn$)
without resorting to self-consistency in $G$ and $W$.\cite{SC}

While the LDA+$\GnWn$ approach is now almost routinely applied to bulk materials,
\cite{Aulbur/Jonsson/Wilkins:2000,Onida/Reining/Rubio:2002,Rinke/etal:2005} $\GnWn$
calculations for GaN and InN have been hampered by the deficiencies of the LDA. For zinc-blende GaN
the LDA+$\GnWn$ band gap of 2.88~eV \cite{Rohlfing/Pollmann:1995,Rohlfing/Pollmann:1998} is still
too low compared to the experimental 3.3~eV, \cite{sitar1992,okumura1994,ramirez-flores1994} while
for InN the LDA predicts a metallic ground state with incorrect band ordering. 
A single \GnWn iteration proves not to be sufficient to restore a proper semiconducting 
state and only opens the band gap to 0.02 - 0.05\,eV, 
\cite{Usuda/Hamada/etal:2004,Kotani/Schilfgaarde:2002} which is still far from the
experimental value of $\sim$0.7\,eV.\cite{Davydov/etal:2002,Wu/etal:2002_1,Nanishi/etal:2003}

Here we apply the $G_0W_0$ approach to DFT calculations in the
exact-exchange optimized effective potential approach (OEPx or OEPx(cLDA) if LDA correlation is included). 
In contrast to LDA and GGA
the OEPx approach is fully self-interaction free and correctly predicts InN to be
semiconducting with the right band ordering in the wurtzite 
phase \cite{Qteish/etal:2005,Rinke/etal:2006}     
as Fig.~\ref{fig:InN_BS} demonstrates.
For II-VI compounds, GaN and ScN we have previously illustrated the mechanism behind
this band gap opening in the OEPx approach, 
\cite{Rinke/etal:2005,Rinke/etal:2006,Qteish/etal:2006,Rinke/psi-k:2007,Rinke/pssb:2008}
which brings the OEPx Kohn-Sham band gap much closer to the experimental one.
Combining OEPx with \GnWn \cite{GW_SI} in this fashion
yields band gaps for II-VI compounds, Ge, GaN and ScN in very good agreement with
experiment.\cite{Rinke/etal:2005,Rinke/etal:2006,Qteish/etal:2006,Rinke/psi-k:2007,Rinke/pssb:2008} 
For wurtzite InN the band gap of 0.7\,eV and the non-parabolicity of the conduction band (CB)
\cite{Rinke/etal:2006} (shown in Fig.~\ref{fig:InN_BS}) strongly supports the recent experimental findings.
\cite{Davydov/etal:2002,Wu/etal:2002_1,Nanishi/etal:2003, Wu/etal:2002,Walukiewicz/etal:2004}
In addition we have shown that 
the source for the startling, wide interval of experimentally observed band gaps can be
consistently explained by the 
Burstein-Moss effect (apparent band gap increase with increasing electron
concentration in the conduction band) by extending our calculations to finite carrier
concentrations. \cite{Rinke/etal:2006}

  \subsection{\label{sec:CA} Computational Parameters}

\begin{table}
  \begin{ruledtabular}
    \begin{tabular}{l|dddd}
        & \multicolumn{1}{c}{$a_0$ [\AA]} & 
          \multicolumn{1}{c}{$c_0$}[\AA] & 
          \multicolumn{1}{c}{$c_0/a_0$} & 
          \multicolumn{1}{c}{$u$}  \\
      \hline
      zb-AlN & 4.370 &      & &      \\
      zb-GaN & 4.500 &      & &      \\
      zb-InN & 4.980 &      & &       \\ \hline
      wz-AlN & 3.110 & 4.980 & 1.6013 &0.382  \\
      wz-GaN & 3.190 & 5.189 & 1.6266 &0.377  \\
      wz-InN & 3.540 & 5.706 & 1.6120 &0.380  \\
      \hline
    \end{tabular} 
  \end{ruledtabular}
  \caption{\label{tab:lat_cut}\small Experimental lattice parameters ($a_0$, $c_0$ and $u$)
           adopted in this work (see text), for 
	   zinc blende (zb) and wurtzite (wz) AlN, GaN, and InN.\cite{footnote_latpar1} }
\end{table}

The LDA and OEPx calculations in the present work were performed with the plane-wave,
pseudopotential code \texttt{S/PHI/nX},\cite{SFHIngX} while for the $G_0W_0$ calculations 
we have employed the $G_0W_0$ space-time method
\cite{Rojas/Godby/Needs:1995} in the \texttt{gwst} implementation.
\cite{GW_space-time_method:1998,GW_space-time_method_enh:2000,GW_space-time_method_surf:2007} 
Local LDA correlation is added in all OEPx calculations.
Here we follow the parametrization of Perdew and Zunger
\cite{Perdew/Zunger:1981} for 
the correlation energy density of the 
homogeneous electron gas based on the data of Ceperley and Alder.\cite{Ceperley/Alder:1980}  
This combination will in the following be denoted OEPx(cLDA). 
Consistent pseudopotentials were used throughout, i.e.\ 
exact-exchange pseudopotentials \cite{Moukara/etal:2000} 
for the OEPx(cLDA) and LDA ones for the LDA calculations.
The cation $d$-electrons were  included 
explicitly.\cite{Qteish/etal:2005,Rinke/etal:2005} 
For additional technical details and convergence
parameters we refer to previous work. \cite{Rinke/etal:2005,Rinke/etal:2006}

  \subsection{\label{sec:latpar} Lattice Parameters}

All calculations are carried out at the experimental lattice constants reported in 
Tab.~\ref{tab:lat_cut} and not at the {\it ab initio} ones
to avoid artificial strain effects in the derived parameter sets.
For an {\it ab initio} determination of the lattic constants consistent with the \GnWn 
calculations the crystal structure would have to be optimized within the \GnWn formalism, too. 
However, \GnWn
total energy calculations for realistic systems have, to our konwledge, not been performed, yet, 
and the quality of the \GnWn total energy for bulk semiconductors has not been assessed 
so far. \cite{Morris/etal}
The alternative {\it ab initio} choices, LDA and OEPx(cLDA), give different lattice constants
\cite{Staedele/etal:1999} and would thus introduce uncontrolable variations in the calculated band
parameters that would aggrevate a direct comparison.

The thermodynamically stable phase of InN at the usual growth conditions is the wurtzite
phase. Reports of a successful growth of the zinc-blende phase have been scarce.
Recently, high-quality films of zb-InN grown on indium oxide have been 
obtained by Lozano \textsl{et al.}\cite{Lozano/etal:2007}  We adopt their lattice constant of 4.98\,\AA,
\cite{Lozano/etal:2007} which is in good agreement with previous reports of 
4.98\,\AA,\cite{Chandrasekhar/etal:1995} 4.986\,\AA,\cite{Cimalla/etal:2003}
and 5.04\,\AA \cite{Lima/etal:2003} for wz-InN grown on different substrates.  For zb-AlN reports of successful growth are even scarcer. Petrov {\it et al.} first achieved to grow AlN in the zinc-blende phase and reported a lattice constant of
$a_0$=4.38~\AA.\cite{Petrov/etal:1992}  This was later refined by Thompson {\it et al.}
to $a$=4.37~\AA,\cite{Thompson/etal:2001} which is the value we adopt in this work.  For zb-GaN we follow the work of Lei {\it et al}.\cite{Lei/etal:1991,Lei/etal:1992} and chose $a_0$=4.50~\AA.

Although wurtzite is the phase predominantly grown for InN, reported values for the
structural parameters still scatter appreciably.\cite{Maleyre/etal:2004} In order to
determine the effect of the lattice constants on the band gap ($E_\tn{g}$) and the 
crystal-field splitting ($\Delta_\tn{CR}$) we have explored the range between the 
maximum and minimum values of $a_0$ and $c_0/a_0$ reported in 
Ref.~\onlinecite{Maleyre/etal:2004} by performing \gw calculations at the values listed 
in Tab.~\ref{tab:cr}. Since $u$ remains 
undetermined in Ref.~\onlinecite{Maleyre/etal:2004} we have optimized it in the LDA. 
Neither $u$, $E_\tn{g}$, nor $\Delta_\tn{CR}$ depends sensitively on the lattice constants 
in this regime and we have therefore adopted the mean values of $a_0$=3.54~\AA, 
$c_0$=5.706~\AA ($c_0/a_0$=1.612) and $u$=0.380 (the LDA-optimized value) for the remainder of this article. 

\begin{table*}
  \begin{ruledtabular}
     \begin{tabular}{ddddddd}
	& & & \mcs{2}{c}{\gw} &	\mcs{2}{c}{OEPx(cLDA)}  \\
	\mcc{$a_0$ (\AA)}    & \mcc{$c_0/a_0$}   & \mcc{u} & \mcc{$\Delta_\tn{CR}$ (eV)} &
	\mcc{$E_\tn{g}$ (eV)} & \mcc{$\Delta_\tn{CR}$ (eV)} &	\mcc{$E_\tn{g}$ (eV)} \\
        \hline   
        3.535 & 1.612 & 0.380 & 0.067 & 0.71 & 0.079 & 1.01 \\
        3.540 & 1.612 & 0.380 & 0.066 & 0.69 & 0.079 & 1.00 \\
        3.545 & 1.612 & 0.380 & 0.065 & 0.68 & 0.079 & 0.98 \\ 
	\hline                                              
%
        3.540 & 1.611  & 0.380 & 0.065 & 0.70 & 0.078 & 1.00 \\
        3.540 & 1.612  & 0.380 & 0.066 & 0.69 & 0.079 & 1.00 \\
        3.540 & 1.613  & 0.380 & 0.068 & 0.69 & 0.081 & 0.99 \\
     \end{tabular}
  \end{ruledtabular}
  \caption{\label{tab:cr}
          Band gap ($E_\tn{g}$), and crystal field splitting ($\Delta_\tn{CR}$) for wurtzite
	  InN in the range of the experimentally reported values of the structural
	  parameters $a_0$ and $c_0/a_0$ (Ref. \onlinecite{Maleyre/etal:2004}) and 
	  $u$ determined in LDA for $a_0$=3.54~\AA and $c_0/a_0$=1.612.
  }
\end{table*}

For wz-AlN and wz-GaN the lattice constants are more established.\cite{Vurgaftman/Meyer:2003,Strite/Morkoc:1992} 
For wz-AlN we adopt Schulz and Thiermann's values of $a$=3.110~\AA, $c$=4.980~\AA, 
and $u$=0.382,\cite{Schulz/Thiermann:1977} which are close to those reported by 
Yim {\it et al. \cite{Yim/etal:1972}}  Schulz and Thiermann also provide a value for the internal 
parameter $u$, which is identical to the one we obtain by relaxing $u$ in the LDA at the 
experimental $a_0$ and $c_0$ parameters. The same is true for wz-GaN.  
Schulz and Thiermann's values of $a$=3.190~\AA and $c$=5.189~\AA 
\cite{Schulz/Thiermann:1977} are close to those first reported by Maruska and 
Tietjen,\cite{Maruska/Tietjen:1969} but in addition offer a value of $u$=0.377, 
which corresponds to our LDA-relaxed value at the same lattice parameters.  
Note, that the lattice parameters of wz-InN and wz-GaN have been refined 
compared to our recently published calculations.\cite{Rinke/etal:2006}  
The influence of the adjustment on the different band parameters will be 
discussed where necessary.

\section{\label{sec:qpe_bs} Band Gaps, Crystal-field Splittings, and Band-gap Deformation Potentials}
We will now discuss the quasiparticle band structure of AlN, GaN and InN in 
their zinc-blende and wurtzite
phases in terms of certain key band parameters such as the band gap ($E_\tn{g}$), the crystal
field splitting ($\Delta_\tn{CR}$) in the wurtzite phase  and the band-gap volume deformation
potentials $\alpha_V$. At the end of this section we will draw a comparison between LDA and
OEPx(cLDA) based \GnWn calculations for AlN.

\subsection{Band Gaps}
\label{sec:gap}

\begin{table*}
  \begin{ruledtabular}
     \begin{tabular}{lldddd}
                  &   param. 			& \mcc{\gw} 		& \mcc{Exp.}           	& \mcc{LDA} 	& \mcc{OEPx(cLDA)} 	\\
                  &    		        & \mcc{(this work)}	&		       	&       
		  \multicolumn{2}{c}{(this work, for comparison)}	                \\
        \hline
 	\multicolumn{6}{c}{wurtzite}\\
        \hline
wz-AlN & $E_\tn{g}$	          	&  6.47 		&\mcc{6.0-6.3\f{2}}     &  4.29  	&  5.73  		\\
       & $\Delta_\tn{CR}$                   	& -0.295 		& \mcc{-0.230\f{3}}     & -0.225 	& -0.334 		\\
wz-GaN & $E_\tn{g}$	 		&  3.24 		& \mcc{3.5\f{5}} 	&  1.78  	&  3.15  		\\
       & $\Delta_\tn{CR}$          		&  0.034 		& \mcc{0.009-0.038\f{5}}&  0.049 	&  0.002 		\\
wz-InN & $E_\tn{g}$	 		&  0.69  		& \mcc{0.65-0.8\f{7}}   & 		&  1.00  		\\
       & $\Delta_\tn{CR}$       		&  0.066 		& \mcc{0.019-0.024\f{8}}& 		&  0.079 		\\
        \hline
 	\multicolumn{6}{c}{zinc blende}\\
        \hline
zb-AlN & $E_\tn{g}^{\Gamma-\Gamma}$ & 6.53 			&		  	& 4.29 		& 5.77 			\\
       & $E_\tn{g}^{\Gamma-X}$ 	& 5.63 			& \mcc{5.34\f{1}}   	& 3.28 		& 5.09 			\\
zb-GaN & $E_\tn{g}$	 		&  3.07 		& \mcc{3.3\f{4}}        &  1.64  	&  2.88  		\\
zb-InN & $E_\tn{g}$	 		&  0.53  		& \mcc{0.6\f{6}}        & 		&  0.81  		\\
     \end{tabular}
  \end{ruledtabular}
    \footnotetext[1]{Reference~\onlinecite{Thompson/etal:2001}}
    \footnotetext[2]{References~\onlinecite{Yim/etal:1972}
                                \onlinecite{chen2004}, \onlinecite{perry1978}, 
                                \onlinecite{guo1994},
                                \onlinecite{Akamaru/etal:2002},
                                \onlinecite{Chen/etal:2004},
				and \onlinecite{Li/etal_AlN:2003}}
    \footnotetext[3]{Reference~\onlinecite{chen2004}.}
    \footnotetext[4]{References~\onlinecite{sitar1992}, \onlinecite{okumura1994}, and \onlinecite{ramirez-flores1994}}
    \footnotetext[5]{Reference~\onlinecite{Vurgaftman/Meyer:2003} and references therein.}
    \footnotetext[6]{Reference~\onlinecite{Schoermann/etal:2006}}    
    \footnotetext[7]{References~\onlinecite{Wu/etal:2002_1}, 
                                \onlinecite{Bhattacharyya/etal:2006},
				\onlinecite{Higashiwaki/etal:2003},
				\onlinecite{Nanishi/etal:2003},
				\onlinecite{Losurdo/etal:2006}, 
				\onlinecite{Davydov/etal:2002}, and \onlinecite{goldhahn2006}}
    \footnotetext[8]{Reference~\onlinecite{goldhahn2006}}    
  \caption{\label{tab:gaps}
          Band gaps ($E_\tn{g}$) and crystal-field splittings ($\Delta_\tn{CR}$)\cite{footnote_dcr} for the wurtzite and zinc-blende phases of AlN, GaN, and InN. All values are given in eV.}
\end{table*}

The \gw band gaps for the three materials and two phases are reported in
Tab.~\ref{tab:gaps} together with the LDA and OEPx(cLDA) values for comparison.
For GaN and InN the \gw band gaps have been 
reported previously in Ref.~\onlinecite{Rinke/etal:2006}.  There we have also argued that 
the wide interval of experimentally observed band gaps for InN can be 
consistently explained by the Burstein-Moss effect.  
The \gw value  
of $0.69$\,eV for wz-InN \cite{footnote_latpar2} supports recent observations of a 
band gap at the lower end of the experimentally reported range.
For zinc-blende InN, which has been explored far less experimentally,
our calculated band gap of 0.53~eV also agrees very well with the recently
measured (and Burstein-Moss corrected) 
0.6~eV. \cite{Schoermann/etal:2006}

For GaN the band gaps of both phases are well established experimentally and our \gw 
calculated values of $3.24$\,eV \cite{footnote_latpar3} and $3.07$\,eV agree to within $0.3$\,eV.

For AlN experimental results for the band gap of the wurtzite phase scatter 
appreciable, whereas for zinc blende only one value has 
-- to the best of our knowledge -- been reported so far. 
Contrary to GaN, the \gw gaps for AlN are larger than the experimentally reported 
values.

\subsection{Crystal-Field Splitting}
\label{sec:delta_cr}
Experimental values for the crystal-field splitting, $\Delta_\tn{CR}$, of wz-GaN 
scatter between 0.009 and 0.038\,eV (Tab.~\ref{tab:gaps}).  The \gw value 
of $0.033$\,eV supports a crystal-field splitting within this range.

Theoretical\cite{carrier2005} and experimental\cite{chen2004} investigations of wz-AlN 
agree upon the fact that the crystal-field splitting of AlN is negative.  Our 
calculations also yield a negative value of $\Delta_\tn{CR}=-0.295$\,eV.  This 
result supports a crystal-field splitting in AlN below $-0.2$\,eV, as reported by 
Chen~\textsl{et al.},\cite{chen2004} rather than a small negative value between 
$-0.01$ and $-0.02$, as implied by the results of 
Freitas~\textsl{et al.}\cite{freitas2003}  For wz-InN a crystal-field splitting 
between $0.019$ and $0.024$\,eV has been reported recently.\cite{goldhahn2006}  This 
value is significantly smaller than the \gw value of $0.07$\,eV. 

The crystal-field splitting is known to be sensitive to lattice deformations, such as 
changes in the $c_0$/$a_0$ ratio or the internal lattice 
parameter $u$.\cite{gil1995,gil1996,wei1996}  Therefore, the discrepancy between experiment 
and theory might stem from the uncertainties of the lattice parameters of wz-InN 
(cf Sec.~\ref{sec:latpar}).  However, varying the $c_0$/$a_0$ ratio or the unit cell volume 
within the experimental range discussed in Section \ref{sec:latpar}  yields values for 
$\Delta_\tn{CR}$ which are always larger than $0.06$\,eV (Tab.~\ref{tab:cr}), leaving only the 
internal lattice parameter $u$ as possible source of error.  This 
parameter is -- at least for GaN -- known to have a large influence on the 
crystal-field splitting.\cite{wei1996} Although the LDA-optimized $u$ values are in 
very good agreement with experimental values for GaN and AlN, experimental 
confirmation of the $u$ parameter of InN is still pending.  We therefore 
calculated the crystal-field splitting of wz-InN for different values of $u$ 
(and $a_0$ and $c_0$ fixed at the values listed in Tab.~\ref{tab:lat_cut}) 
between $0.377$ and $0.383$.  
Generally, $\Delta_\tn{CR}$ decreases with increasing $u$, but  even for $u$ as large as
$0.383$, the crystal-field splitting is still larger than $0.05$\,eV 
The discrepancy between the experimental report and the \gw calculations can hence not be 
attributed to the uncertainties in the lattice parameters and has to remain unsettled 
for the time being.   

\subsection{Band Gap Deformation Potentials}
\label{sec:defpot}
\begin{table}
  \begin{ruledtabular}
     \begin{tabular}{llddd}
                  &    & \mcc{\gw} & \mcc{LDA} & \mcc{OEPx(cLDA)} \\
		  &    & \mcc{(this work)}     &   
		         \multicolumn{2}{c}{(this work, for compar.)}               \\
        \hline   
	\multicolumn{5}{c}{wurtzite}\\
        \hline   
	AlN  & $\alpha_{V}^{\rm wz}$	          	& -9.8 	& -8.8  & -8.9  \\ 	
	GaN  & $\alpha_{V}^{\rm wz}$	 		& -7.6  & -6.8  & -6.5  \\ 
	InN  & $\alpha_{V}^{\rm wz}$	 		& -4.2  & 	& -3.0   \\ 
        \hline   
	\multicolumn{5}{c}{zinc blende}\\
        \hline   
	AlN  & $\alpha_{V}^{{\rm zb}\:\Gamma-\Gamma}$& -10.0 & -9.1 	& -9.1 	\\ 		
	     & $\alpha_{V}^{{\rm zb}\:\Gamma-X}$     &  -1.8 & -0.6 	& -0.9 	\\
	GaN  & $\alpha_{\Gamma}^{\rm zb}$	 		& -7.3  & -6.4  & -6.1  \\ 
	InN  & $\alpha_{V}^{\rm zb}$	 		& -3.8  & 	& -2.6   \\
     \end{tabular}
  \end{ruledtabular}
  \caption{\label{tab:defpots}
          Volume band gap deformation potentials ($\alpha_{V}$) for the wurtzite and zinc-blende 
	  phases of AlN, GaN, and InN. The volume deformation potentials can be transformed into 
	  pressure deformation potentials using the bulk moduli of the respective materials 
	  (see text). All values are given in eV.}
\end{table}
For the hydrostatic band gap deformation potentials the band gaps have been
calculated at different volumes ($V$) between $\pm 2$\,\% around the 
equilibrium volume $V_0$. 
In the explored volume range the band gaps vary linearly with $\ln{(V/V_0)}$. The linear coefficient is then taken as the hydrostatic volume deformation potential $\alpha_V$. 
The calculated band gap deformation potentials are listed in Tab.~\ref{tab:defpots} for
LDA, OEPx(cLDA), and \gw. We observe that for all compounds and phases the
quasiparticle deformation potential is larger in magnitude than that of the DFT 
(LDA and OEPx(cLDA)) calculations, i.e., the band gaps vary stronger with volume 
deformations. 
Hydrostatic band gap deformation potentials obtained from LDA+U calculations have recently
been reported for the wurtzite phases of GaN and InN \cite{Janotti/VandeWalle:2007}. 
Unlike in OEPx, where an improved 
description of the $p$-$d$ hybridization is achieved by the full removal of the 
self-interaction for all valence states, the LDA+U approach reduces the $p$-$d$ repulsion 
by adding an on-site Coulomb correlation U only to the semicore $d$-electrons. 
With reference to the \gw deformation potentials, the LDA+U improves upon LDA for GaN
($\alpha_{V}^{{\rm LDA+U}}$=-7.7~eV), but worsens
for InN ($\alpha_{V}^{{\rm LDA+U}}$=-3.1~eV, $\alpha_{V}^{{\rm LDA}}$=-4.2~eV 
in Ref. \onlinecite{Janotti/VandeWalle:2007}).

Experimentally the band gap deformation potential is usually measured as a function of the 
applied pressure, which aggravates a direct comparison to our calculated volume deformation 
potentials. However, since $B=-dP/d\ln{V}$, where $B$ is the bulk modulus and $P$ the 
pressure, the pressure deformation potential $\alpha_P$ can be expressed in terms of 
$\alpha_V$ according to $\alpha_P=-\alpha_V/B$. 

Experimentally reported values for the bulk modulus of wz-GaN scatter between 1880 and 2450 
kbar.\cite{Schulz/Thiemann:1977,
      Xia/etal:1993,
      Ueno/etal:1994,
      Landolt-Boernstein_1,
      Perlin/etal:1992}  Using these values, our volume deformation potential of 
$\alpha_V$=\mbox{-7.6}~eV would translates into a pressure deformation potential in the 
range of 3.1 - 4.0~meV/kbar, which is comparable to the experimentally determined range of 
3.7 and 4.7~meV/kbar. 
\cite{Camphausen/Connell:1971,
      Perlin/etal:1992,
      Shan/etal:1995,
      Kim/etal:1995,
      Perlin/etal:1999,
      Reimann/etal:1998}
This large uncertainty has been partially ascribed to the low quality of earlier samples 
and substrate-induced strain effects. \cite{Perlin/etal:1999}
The fact that the pressure dependence of the band gap 
is sublinear (unlike the volume dependence) further questions the accuracy of linear or 
quadratic fits for the extraction of 
the deformation potentials in the experiments.\cite{Perlin/etal:1999} 

For wz-InN experimentally reported values are sparse. Franssen {\it et al.}
determined a hydrostatic pressure deformation potential of
2.2~meV/kbar\cite{Franssen/etal:2006}, while Li {\it et al.} found
3.0~meV/kbar.\cite{Li/etal:2003} This range agrees with our theoretical one 
of 2.8 - 3.3~meV/kbar, using for the conversion of volume to pressure
deformation potentials the bulk modulus range of 1260 -
1480 kbar \cite{Ueno/etal:1994,Landolt-Boernstein_1} quoted in the literature. 

For wz-AlN we are only aware of one experimental study reporting a pressure
deformation potential of 4.9~meV/kbar.\cite{Akamaru/etal:2002}
With experimental bulk moduli between 1850 and 2079\,kbar 
\cite{Xia/etal:1993_2,Mashimo/etal:1999,Ueno/etal:1992}
 the \gw pressure deformation potential of wz-AlN would fall between 4.7 and 5.3~meV/kbar straddling
the experimentally reported value.

To our knowledge, no experimental information on the deformation potential
of zb-AlN and zb-InN are available. For zb-GaN our computed volume deformation potential of $\alpha_V$=\mbox{-7.3}~eV
translates to a pressure deformation potential range of 3.0 - 3.9~meV/kbar using the same bulk modulus range as for wz-GaN. This range is slightly below the experimentally reported range of 4.0 - 4.6~meV/kbar.
\cite{Hwang/etal:1994,Reimann/etal:1998,Liu/etal:1999}
Employing a semi-empirical approach to overcome the band gap underestimation of the
LDA (LDA-plus-correction (LDA+C), see also discussion in Section \ref{sec:comp_exp}), 
Wei and Zunger found volume deformation potentials of $-10.2$\,eV
[zb-AlN ($\Gamma-\Gamma$)], $-1.1$\,eV [zb-AlN ($\Gamma-X$)], $-7.4$\,eV (zb-GaN) 
and $-3.7$\,eV (zb-InN)\cite{Wei/Zunger:1999} in good agreement with our full \gw 
calculations (see Tab.~\ref{tab:defpots}).

\subsection{Comparison between LDA+\textit{G$_0$W$_0$} and 
           OEPx(cLDA)+\textit{G$_0$W$_0$}}
\label{sec:comp_GWs}

\begin{figure}[t] 
   \epsfig{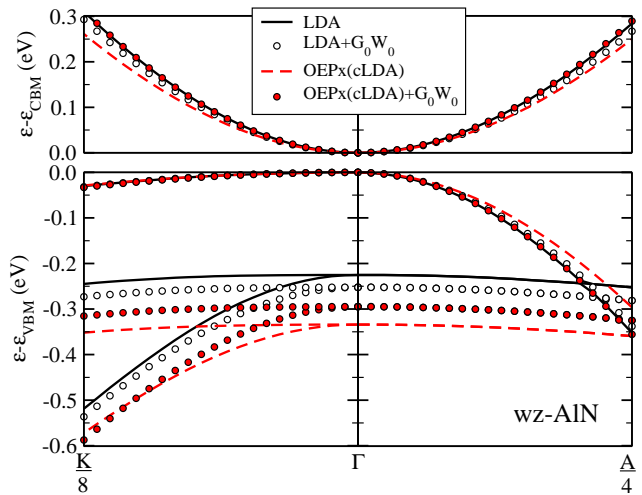}
   \caption{\label{fig:AlN_BS}
	    Comparison between LDA (solid lines), LDA+$\GnWn$ (open circles), OEPx(cLDA)
	    (dashed lines) and \gw (solid circles) for wz-AlN. 
	    In both $\GnWn$ calculations bands are essentially shifted rigidly compared to
	    the LDA, whereas OEPx(cLDA) yields bands with different dispersion.}
\end{figure}

For the materials presented in this article a meaningful comparison between LDA and OEPx(cLDA)
based \GnWn calculations can only be constructed for AlN 
for reasons given in Section \ref{sec:OEPx_GW}. 
Figure~\ref{fig:AlN_BS} displays the band structure of wz-AlN in the four
approaches discussed in this
Article. 
The ``band gap problem'' 
has been eliminated from this comparison by aligning the conduction bands at the minimum of
the lowest conduction band ($\epsilon_{\rm CBM}$) and the valence bands at the maximum of the 
hightest valence band ($\epsilon_{\rm VBM}$).
For this large gap material three main conclusions can be drawn from Fig.~\ref{fig:AlN_BS}.
First, LDA and both $\GnWn$ calculations yield very similar band dispersions. 
Or in other words
the $\GnWn$ corrections to the LDA in the LDA+$\GnWn$ approach are not $\bfk$-point dependent shifting
bands almost rigidly. A rigid shift between conduction and valence bands is frequently
referred to as ``scissor operator''. Fig.~\ref{fig:AlN_BS}, however, illustrates that this
shift is not identical for all bands, which cannot be attributed to a 
single scissor operator. 
Second, the dispersion obtained in OEPx(cLDA) deviates from the other three approaches,
which is consistent with the observation made for wz-InN in Fig.~\ref{fig:InN_BS}.
We attribute this behaviour to the approximate treatment of 
correlation in the OEPx(cLDA) approach and the fact that the band
structure in OEPx(cLDA) is a Kohn-Sham and not a quasiparticle band
structure. While the LDA benefits from a fortuitous
error cancellation between the exchange and the correlation part
\cite{Gunnarsson/Lundqvist:1976}, this is
no longer the case once exchange is treated exactly in the OEPx(cLDA)
scheme. Using a quasiparticle approach with a more sophisticated description of
correlation, like the $GW$ method, then notably
changes the dispersion of the OEPx(cLDA) bands. As we will demonstrate in the next Section
this will lead to markedly different band parameters 
not only for the conduction but also for the valence bands 
(cf Tab.~\ref{tab:exp}). 
Against common believe OEPx(cLDA) calculations without
subsequent $\GnWn$ calculations may therefore provide a distorted picture and we would
advise against deriving band parameters from OEPx
or OEPx(cLDA) band structures alone.
Third, unlike in the LDA+$\GnWn$ case the $\GnWn$ corrections to the OEPx(cLDA) starting
point become $\bfk$-point dependent, a fact already observed for GaN and II-VI compounds.
\cite{Rinke/etal:2005} Most remarkably and in contrast to what we observe for GaN and InN 
(see Sec. \ref{sec:other_par}) the corrections are such that the band
dispersion now agrees again with that obtained from the LDA and the LDA+$\GnWn$ approach. 
Note also
that  both the band gap and the crystal field splitting still differ slightly 
between LDA+$\GnWn$ and \gw for AlN 
($E_g$:            LDA+\GnWn: 5.95~eV, \gw: 6.47~eV, 
 $\Delta_\tn{CR}$: LDA+\GnWn: -0.252~eV, \gw: -0.295~eV).
Unlike for GaN and InN, experimental uncertainties do, at present, not permit a rigorous 
assessment, which of the two \GnWn calculations provides a better description for AlN (see
also Sections \ref{sec:gap} and \ref{sec:delta_cr}).

\section{\label{sec:kp_bs}Band Dispersion Parameters}
        {%

We will now turn our attention to band parameters that describe the band 
dispersion in the vicinity of the $\Gamma$-point: the effective masses, the 
Luttinger(-like) parameters, and the $E_\tn{P}$ parameters.  These parameters 
are obtained by means of the $\kpe$-method. The $\kpe$-method is a 
well-established approach that permits a description of semiconductor band 
structures in terms of parameters that can be accessed experimentally.  
Throughout this paper, we use four-band $\kpe$-theory, which is typically used 
to describe direct-gap materials, mostly in its spin-polarized form as 
eight-band $\kpe$-theory.  The $\kpe$ Hamiltonian and all relevant formulas 
are given in Appendix \ref{sec:kp_ham}.
The $\kpe$-method is a widely accepted technique for, e.g., the interpretation 
of experimental data \cite{Wu/etal:2002, fu2004} or modeling of semiconductor 
nanostructures and \mbox{(opto-)}electronic devices.\cite{gershoni,stier1999,andreev2000,fonoberov2003,winkelnkemper2006,hackenbuchner2001,lee2006} Its accuracy, however, depends crucially on the quality of the input 
band parameters, like effective electron masses, Luttinger-parameters, etc., 
which have to be derived either experimentally or from band structure calculations.  As alluded to in the introduction, many important band parameters of the group-III nitrides GaN, InN, and AlN are still unknown.  In particular, the band structure of InN is currently the subject of active research in both experiment and theory.

In this paper, we use the $\kpe$-method to derive band dispersion parameters 
from \gw band structures.  This approach has certain advantages over a 
simple parabolic approximation around the $\Gamma$ point.  First, the $\kpe$ 
band structure is valid, not only directly at the $\Gamma$-point, but also in 
a certain $\mathbf{k}$-range around it.  This allows to extend the fit to 
larger $\mathbf{k}$'s and thereby increases the accuracy of the fitted parameters.  
Second, the $\kpe$-method is capable of describing non-parabolic bands, such 
as the CB of InN\cite{Wu/etal:2002,Rinke/etal:2006} and can therefore also be applied to 
accurately determine values for the effective electron masses and $E_\tn{P}$ 
parameters in InN. 

\subsection{Computational details}
For an accurate fit of the $\kpe$ parameters to the quasiparticle band 
structure a small reciprocal lattice vector spacing is required. Since 
most $GW$ implementations evaluate the self-energy $\Sigma$ (the perturbation operator that
links the Kohn-Sham with the quasiparticle system) in 
reciprocal space, the matrix elements with respect to the Kohn-Sham wave functions
$\bracket{\phi_{\tn{n}\bfq}}{\Sigma(\epsilon_{\tn{n}\bfq}^{\tn{qp}})}{\phi_{\tn{n}\bfq}}$ 
required for the quasiparticle corrections are only available on the 
$\bfk$-points of the underlying $\bfk$-grid. A fine sampling of the 
$\Gamma$-point region would therefore be equivalent to using formidably 
large $\bfk$-grids in the computation. Most interpolation schemes that are 
frequently employed to calculate the quasiparticle corrections for arbitrary 
band structure points are to no avail in this case, because they do not add 
new information to the fitting problem at hand. Existing schemes to directly 
compute the self-energy for band structure $\bfq$-points not contained in 
the $\bfk$-grid (see, for instance, 
Ref.~\onlinecite{Aulbur/Jonsson/Wilkins:2000} for an overview) are usually not 
implemented.

In the $GW$ space-time method \cite{Rojas/Godby/Needs:1995} these problems are 
easily circumvented, because the self-energy is computed in real-space 
[$\Sigma_\bfR(\bfr,\bfrp;\epsilon)$]. By means of Fourier interpolation, 
\begin{equation}
\label{eq:FI}
  \Sigma_\bfq(\bfr,\bfrp;\epsilon)
          =\sum_\bfR \Sigma_\bfR(\bfr,\bfrp;\epsilon)
                  e^{-i\bfq \cdot \bfR}\quad,
\end{equation}
the self-energy operator can be calculated at arbitrary $\bfq$-points. 
\cite{GW_space-time_method:1998}
The matrix elements 
$\bracket{\phi_{\tn{n}\bfq}}{\Sigma_\bfq(\epsilon)}{\phi_{\tn{n}\bfq}}$ 
are then obtained by integration
over $\bfr$ and $\bfrp$. In this fashion the relevant Brillouin zone regions 
for the band structure fitting can be calculated efficiently without compromising accuracy.

The $\kpe$ Hamiltonian and all parameter relations are given in 
Appendices \ref{sec:kp_ham} and \ref{sec:meff}. 
To determine the $\kpe$ Hamiltonian for a given band structure with band gap $E_\tn{g}$ and crystal-field splitting $\Delta_\tn{CR}$ we fit the 
	parameters $m_\tn{e}^\tn{i}$, $A_\tn{i}$, $\gamma_\tn{i}$, and 
	$E_\tn{P}^\tn{i}$.  This is achieved by least-square-root fitting of 
	the $\kpe$ band structure to the \gw band structure in the vicinity of 
	$\Gamma$.  For the wurtzite phases the directions 
	$\Sigma$, $\Lambda$, $\mathrm{T}$, and $\Delta$ have been included in 
	the fit, represented by 22 equidistant $\mathbf{k}$-points from 
	$\Gamma$ to $\frac{\tn{M}}{8}$ ($\frac{\tn{L}}{8}$, $\frac{\tn{K}}{8}$)
	and 22 equidistant points from $\Gamma$ to $\frac{\tn{A}}{4}$. For 
	the zinc-blende phases the directions $\Sigma$, $\Delta$, and $\Lambda$ have 
	been included, each with 22 $\mathbf{k}$-points from $\Gamma$ to 
	$\frac{\tn{K}}{8}$ ($\frac{\tn{X}}{8}$, $\frac{\tn{L}}{8}$).

\subsection{\label{sec:kp_dis}Band parameters of GaN, AlN, and InN}
        \begin{figure*}[t]
                \includegraphics[width=0.8\textwidth,clip]{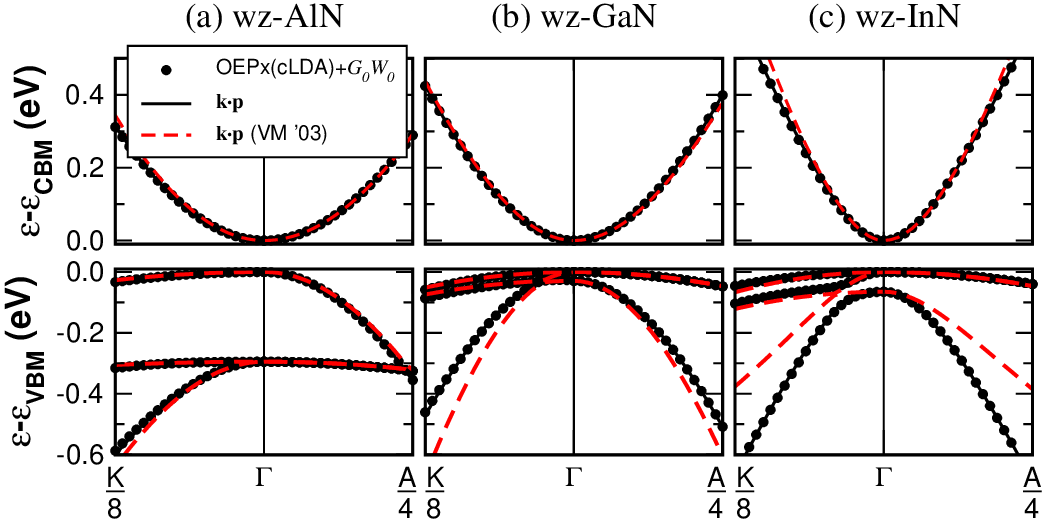}
                \caption{\label{fig:kp_bs_wz}  Band structure wz-AlN (a), wz-GaN (b), and wz-InN (c) 
		in the vicinity of $\Gamma$.  The graphs show the \gw  band structure (black circles), the corresponding $\kpe$ band structure (black solid lines), and the $\kpe$ band structure using the parameters recommended by Vurgaftman and Meyer\cite{Vurgaftman/Meyer:2003,footnote_momme3} (VM '03) (red dashed lines).}
        \end{figure*}
        \begin{figure*}[t]
                 \includegraphics[width=0.8\textwidth,clip]{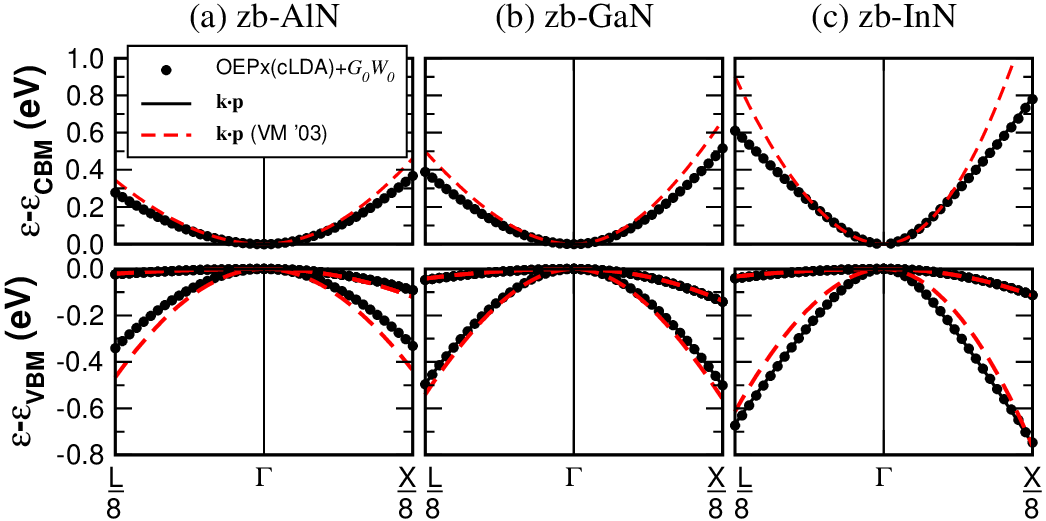}
                 \caption{\label{fig:kp_bs_zb}  Band structure zb-AlN (a), zb-GaN (b), 
		 and zb-InN (c) in the vicinity of $\Gamma$.  The graphs show the \gw  band structure 
		 (black circles), the corresponding $\kpe$ band structure (black solid lines), and the $\kpe$ band structure using the parameters recommended by Vurgaftman and Meyer\cite{Vurgaftman/Meyer:2003,footnote_momme3} (VM '03) (red dashed lines).}
	\end{figure*}
                        \begin{table}
        			\begin{ruledtabular}
                                        \caption{
                                                \label{tab:wz}
                                                Recommended band parameters for the wurtzite phases of GaN, InN, and AlN derived from the \gw band structures.
                                        }
        				\begin{tabular}{lddd}
        param.         		        & \mcc{AlN}     & \mcc{GaN}       & \mcc{InN}               \\
        \hline                                    	 			 		        
        $m_\tn{e}^{\parallel}$	        & 0.322         & 0.186           & 0.065	                    \\
        $m_\tn{e}^{\perp}$	        & 0.329         & 0.209           & 0.068	                    \\
        $A_1$			        & -3.991        & -5.947          & -15.803	            \\
        $A_2$			        & -0.311        & -0.528          & -0.497	            \\
        $A_3$			        & 3.671         & 5.414           & 15.251	            \\
        $A_4$			        & -1.147        & -2.512          & -7.151	            \\
        $A_5$			        & -1.329        & -2.510          & -7.060	            \\
        $A_6$			        & -1.952        & -3.202          & -10.078	            \\
        $A_7$ (eV\AA)		        & 0.026         & 0.046           & 0.175	                    \\
        $E_\tn{P}^{\parallel}$ (eV)     & 16.972        & 17.292          & 8.742	                    \\
        $E_\tn{P}^{\perp}$ (eV)         & 18.165   	& 16.265          & 8.809   	            \\
        
        				\end{tabular}
        			\end{ruledtabular}
        		\end{table}
                        \begin{table}
        			\begin{ruledtabular}
                                        \caption{
                                                \label{tab:zb}
                                                Recommended band parameters for the zinc blende phases of GaN, InN, and AlN derived from the \gw band structures. The effective hole masses for the $HH$ and $LH$ band have been calculated from the Luttinger parameters. (see appendix~\ref{sec:meff})
                                        }
        				\begin{tabular}{lddd}
        param. 		& \mcc{AlN}    & \mcc{GaN} 		& \mcc{InN}\\
        \hline                                                 
        $m_\tn{e}(\Gamma)$      & 0.316	       & 0.193	       & 0.054	       \\
        $\gamma_1$		& 1.450        & 2.506         & 6.817        \\
        $\gamma_2$		& 0.349        & 0.636         & 2.810        \\
        $\gamma_3$		& 0.597        & 0.977         & 3.121        \\
        $m_{hh}^{[001]}$   	& 1.330        & 0.810         & 0.835        \\
        $m_{hh}^{[110]}$   	& 2.634        & 1.384         & 1.368        \\
        $m_{hh}^{[111]}$   	& 3.912        & 1.812         & 1.738        \\
        $m_{lh}^{[001]}$   	& 0.466        & 0.265         & 0.080        \\
        $m_{lh}^{[110]}$   	& 0.397        & 0.233         & 0.078        \\
        $m_{lh}^{[111]}$   	& 0.378	       & 0.224	       & 0.077	       \\
        $E_\tn{P}$ (eV)            & 23.844       & 16.861        & 11.373       \\
        
        				\end{tabular}
        			\end{ruledtabular}
        		\end{table}
The parameters obtained by fitting to the \gw band structures are 
listed in Tab.~\ref{tab:wz} (wz) and Tab.~\ref{tab:zb} (zb).  
The resulting $\kpe$ band structures are plotted in 
Figs.~\ref{fig:kp_bs_wz} and~\ref{fig:kp_bs_zb} (black solid lines) 
together with the respective \gw data (black circles).  The excellent agreement 
of the $\kpe$ and \gw band structures illustrates that the band 
structures of the wurtzite and zinc-blende phases of all three materials are 
accurately described by the $\kpe$-method within the chosen $\mathbf{k}$-ranges.  
Additionally, the $\kpe$ band structures based on the parameters 
recommended by Vurgaftman and Meyer\cite{Vurgaftman/Meyer:2003,footnote_momme3} (VM '03) are 
shown (red dashed lines).  As alluded to in the introduction, their recommendations are 
based on available experimental data and selected theoretical values, 
representing the state-of-the-art parameters up until the year of 
compilation (2003). We will also compare our results to more recent experimentally and 
theoretically derived parameters. (see Tab.~\ref{tab:exp})

In the following we will show that the parameters derived from the \gw 
calculations match all available experimental data to good accuracy.  
A comparison to parameters derived by other, theoretical or 
semi-empirical, methods will be presented thereafter.

Before we proceed, however, 
we would like to emphasize two points regarding the relation between the VB parameters $A_\tn{i}$ 
 and the effective hole masses in wurtzite crystals: (i) Two different 
sets of equations, connecting the effective hole masses to the $A_\tn{i}$ parameters, 
are used in the literature.  Reference~\onlinecite{chuang1996} lists both; 
one is labeled ``Near the band edge ($k\rightarrow 0$)'' and the other 
``Far away from the band edge ($k$ is large)''.  
The latter is widely used to calculate the effective hole 
masses.\cite{suzuki1995, fritsch2003, ren1999,pugh1999-1}  However, the experimentally 
relevant effective masses are those close to $\Gamma$.  Thus, we use the 
``Near the band edge'' equations (see Appendix \ref{sec:meff}) 
throughout this work.  Quoted values differ 
from the original publications in cases where the original work uses the 
``Far away from the band edge'' equations.  (ii) The Luttinger-like 
parameters, $A_\tn{i}$, are independent of the spin-orbit and crystal-field interaction 
parameters $\Delta_\tn{SO}$ and $\Delta_\tn{CR}$; the effective hole masses, 
however, differ for different $\Delta_\tn{SO}$ and $\Delta_\tn{CR}$ parameters. Only the $A$-band ($C$-band in AlN) hole masses can be calculated from the Luttinger-like 
parameters alone.  All other hole masses depend additionally on the choice of the spin-orbit and crystal-field splitting energies.\cite{chuang1996}  Thus, effective $B$- and $C$-band 
($A$- and $B$-band in AlN) hole masses derived from different sets of 
Luttinger-like parameters are comparable, only if the same $\Delta_\tn{SO}$ 
and $\Delta_\tn{CR}$ values are assumed.

	\subsubsection{\label{sec:comp_exp} Comparison to experimental values}
	
                \begingroup
                \squeezetable
                        \begin{table*}
        			\begin{ruledtabular}
                                        \caption{
                                                \label{tab:exp}
                                               Band parameters of wurtzite group-III
					       nitrides: Comparison to parameters from the
					       literature. Listed are experimental
parameters, the parameters recommended by Vurgaftman and Meyer\cite{Vurgaftman/Meyer:2003} (VM
'03), parameters using the LDA and OEPx(cLDA), parameters calculated by Carrier and Wei\cite{carrier2005} (CW '05) using LDA+C, and values determined using the empirical pseudopotential method (EPM) by several different groups.}
        				\begin{tabular}{lldcddddc}
                        & param.                        & \mcc{\gw}       &	\mcc{exp.}	& \mcc{VM '03\f{1}}     & \mcc{LDA}	&	\mcc{OEPx(cLDA)}	& \mcc{LDA+C\f{2}}     & \mcc{EPM}                          \\
			&				& \mcc{(recommended)}		&		
			&			&	\multicolumn{2}{c}{(this work, for comparison)}			&		       &			\\				
\hline                                                                                                                          
\multirow{6}{*}{AlN}    & $m_\tn{e}^{\parallel}$	& 0.32                  & 0.29-0.45\f{3}        & 0.32                  & 0.33		& 0.38		& 0.32                  & 0.23\f{4}, 0.24\f{5},0.27\f{6}                     \\
                        & $m_\tn{e}^{\perp}$		& 0.33                  & 0.29-0.45\f{3}        & 0.30                  & 0.32		& 0.39		& 0.33                  & 0.24\f{4}, 0.25\f{5},0.18\f{6}                     \\
                        & $E_\tn{P}^{\parallel}$ (eV)   & 16.97                 & -                     & -	                & 16.41		& 16.89		& -	                & -	                                                \\
                        & $E_\tn{P}^{\perp}$ (eV)       & 18.17   	        & -	                & -	                & 17.45		& 17.51		& -	                & -       	                                        \\
                        & $m_{C}^{\parallel}$ 	        & 3.13	                & -                     & 3.57                  & 3.46		& 3.72		& 3.43             	& 2.38\f{4}, 1.87-1.95\f{5},2.04\f{6}                 \\
                        & $m_{C}^{\perp}$     	        & 0.69                  & -                     & 0.59                  & 0.68		& 0.83		& 0.68             	& 0.49\f{4}, 0.43-0.48\f{5},0.36\f{6}                 \\
\hline                                                                                                                          
\multirow{10}{*}{GaN}   & $m_\tn{e}^{\parallel}$        & 0.19                  & 0.20\f{7}             & 0.20                  & 0.15		& 0.23		& 0.20                  & 0.14\f{4}, 0.14\f{5},0.16\f{6}                      \\
                        & $m_\tn{e}^{\perp}$	        & 0.21                  & 0.20\f{7}             & 0.20	                & 0.17		& 0.26		& 0.22	                & 0.15\f{4}, 0.15\f{5},0.12\f{6}                      \\
                        & $E_\tn{P}^{\parallel}$ (eV)   & 17.29 	        & 17.8-18.7\f{8}, 19.8\f{9} & -               & 12.93		& 16.14		& -	                & -	                                                \\
                        & $E_\tn{P}^{\perp}$ (eV)       & 16.27                 & 16.9-17.8\f{8}, 19.8\f{9} & -               & 11.30		& 14.07		& -	                & -       	                                        \\
                        & $m_{A}^{\parallel}$ 	        & 1.88  	        & 1.76\f{10}             & 1.89                  & 1.92		& 2.20		& 2.04                  & 1.89\f{11}, 2.37\f{4}, 1.45-1.48\f{5},1.27\f{6}      \\
                        & $m_{A}^{\perp}$     	        & 0.33  	        & 0.35\f{10}             & 0.26                  & 0.26		& 0.38		& 0.39                  & 0.26\f{11}, 0.49\f{4}, 0.26-0.27\f{5},0.20\f{6}      \\
                        & $m_{B}^{\parallel}$ 	        & 0.37\f{12}, 0.92\f{13}& 0.42\f{10}             & -                     & -		& -		& 0.85                  & -                                                     \\
                        & $m_{B}^{\perp}$ 	        & 0.49\f{12}, 0.36\f{13}& 0.51\f{10}             & -                     & -		& -		& 0.43                  & -                                                     \\
                        & $m_{C}^{\parallel}$ 	        & 0.26\f{12}, 0.19\f{13}& 0.30\f{10}             & -                     & -		& -		& 0.19                  & -                                                     \\
                        & $m_{C}^{\perp}$     	        & 0.65\f{12}, 1.27\f{13}& 0.68\f{10}             & -                     & -		& -		& 1.05                  & -                                                     \\
\hline                                                                                                                          
\multirow{6}{*}{InN}    & $m_\tn{e}^{\parallel}$        & 0.065                 & 0.07\f{14}, 0.05\f{15}, 0.04\f{16}, 0.085\f{17} & 0.07                  & -	& 0.11		& 0.06                  & 0.072\f{18}, 0.10\f{5},0.14\f{6}                                \\
                        & $m_\tn{e}^{\perp}$		& 0.068                 & 0.07\f{14}, 0.05\f{15}, 0.05\f{16}, 0.085\f{17} & 0.07                  & -	& 0.12		& 0.07                  & 0.068\f{18}, 0.10\f{5},0.10\f{6}                                \\
                        & $E_\tn{P}^{\parallel}$ (eV)   & 8.74                  & 10\f{14}, 9.7\f{15}    & -	                & -		& 9.22		& -	                & -	                                                \\
                        & $E_\tn{P}^{\perp}$ (eV)       & 8.81   	        & 10\f{14}, 9.7\f{15}    & -	                & -		& 8.67		& -	                & -       	                                        \\
                        & $m_{A}^{\parallel}$           & 1.81	                & -                     & 1.56                  & -		& 2.12		& 2.09                  & 2.56-2.63\f{18}, 1.35-1.43\f{5},1.56\f{6}                                       \\
                        & $m_{A}^{\perp}$            	& 0.13	                & -                     & 0.17                  & -		& 0.21		& 0.14                  & 0.14-0.15\f{18}, 0.18-0.20\f{5},0.17\f{6}                                       \\
        				\end{tabular}
        			\end{ruledtabular}
        			\footnotetext[1]{Vurgaftman and Meyer, Ref.~\onlinecite{Vurgaftman/Meyer:2003}.}
        			\footnotetext[2]{Carrier and Wei, Ref.~\onlinecite{carrier2005}.}
        			\footnotetext[3]{Silveira \textsl{et al.}, Ref.~\onlinecite{silveira2004}.}
        			\footnotetext[4]{Fritsch \textsl{et al.}, Ref.~\onlinecite{fritsch2003}.}
        			\footnotetext[5]{Dugdale \textsl{et al.}, Ref.~\onlinecite{dugdale2000}.}
        			\footnotetext[6]{Pugh \textsl{et al.}, Ref.~\onlinecite{pugh1999-1}.}
        			\footnotetext[7]{Reference~\onlinecite{Vurgaftman/Meyer:2003} and references therein.}
        			\footnotetext[8]{Rodina and Meyer, Ref.~\onlinecite{rodina2001_2}.}
        			\footnotetext[9]{Shokhovets \textsl{et al.}, Ref.~\onlinecite{shokhovets2005}.}
                                \footnotetext[10]{Rodina \textsl{et al.}, Ref.~\onlinecite{rodina2001}.}
                                \footnotetext[11]{Ren \textsl{et al.}, Ref.~\onlinecite{ren1999}.}
        			\footnotetext[12]{Calculated using $\Delta_\tn{SO}=0.019$\,eV and $\Delta_\tn{CR}=0.010$\,eV (Ref.~\onlinecite{rodina2001}).}
        			\footnotetext[13]{Calculated using $\Delta_\tn{SO}=0.016$\,eV and $\Delta_\tn{CR}=0.025$\,eV (Ref.~\onlinecite{carrier2005}).}
        			\footnotetext[14]{Wu \textsl{et al.}, Ref.~\onlinecite{Wu/etal:2002}.}
        			\footnotetext[15]{Fu \textsl{et al.}, Ref.~\onlinecite{fu2004}.}
        			\footnotetext[16]{Hofmann \textsl{et al.}, Ref.~\onlinecite{hofmann2006}.}
        			\footnotetext[17]{Inushima \textsl{et al.}, Ref.~\onlinecite{inushima2003}.}
        			\footnotetext[18]{Fritsch \textsl{et al.}, Ref.~\onlinecite{fritsch2004}.}
        		\end{table*}
                \endgroup
		Experimentally, the band structure of a semiconductor is 
		accessible only indirectly, via band parameters like $E_\tn{g}$,
		$\Delta_\tn{SO}$, $\Delta_\tn{CR}$, and the effective masses.
		Angle resolved direct and inverse photoemission experiments, 
		which would, in principle, directly probe the quasiparticle 
		band structure, are not accurate enough, yet, to determine the 
		band structure with sufficient accuracy.

		The dispersion of the conduction band around the $\Gamma$ point depends 
		only on the effective electron masses and $E_\tn{P}$ parameters,
		which are accessible experimentally.
		The valence-band parameters, $A_\tn{i}$, cannot be obtained directly experimentally, but can be related to the effective hole masses (see appendix~\ref{sec:meff}), which, in turn, can be
		measured.

                The available experimental values for the wurtzite phases 
		are listed in Tab.~\ref{tab:exp}. 
		For the thermodynamically metastable zinc-blende phases of GaN, AlN, and InN hardly any experimental 
		reports on their band dispersion parameters are available so 
		far. Therefore, we restrict the discussion to the wurtzite phases, 
                for which experimental data on, at least, the effective electron masses 
		are available.  For wz-InN also $E_\tn{P}$ has been determined,
		by fitting a simplified $\kpe$-Hamiltonian to the experimental 
		data.\cite{Wu/etal:2002,fu2004} 
		For wz-GaN, values for $E_\tn{P}$\cite{rodina2001_2,shokhovets2005} and also several reports on the effective hole masses are available.\cite{footnote_momme2} 

                \paragraph*{Wurtzite GaN.}  The \gw effective electron masses 
		in wz-GaN ($m_\tn{e}^{\parallel}=0.19\,m_0$, $m_\tn{e}^\perp=0.21\,m_0$) 
		are in very good agreement with experimental values, which 
		scatter around $m_\tn{e}=0.20\,m_0$.\cite{footnote_momme2}  
		However, our calculations predict an anisotropy of the 
		electron masses of about 10~\%, which is larger than values 
		found experimentally. ($<1$\,\% - $6$\,\%\cite{perlin1996,kasic2000,meyer1995})
		Our $E_\tn{P}$ values of $17.3$\,eV and $16.3$\,eV support those obtained by Rodina and Meyer\cite{rodina2001_2} ($\approx 18.3$\,eV and $\approx 17.3$\,eV), rather than a larger value of $E_\tn{P}\approx 19.8$\,eV reported recently by Shokhovets \textsl{et al.}\cite{shokhovets2005} 

		A detailed analysis of the effective hole masses has been presented by 
		Rodina \textsl{et al.}\cite{rodina2001} Note, that only the 
		$A$-band masses in their work have been extracted directly from experimental data. All other 
		effective hole masses have been calculated from the $A$-band 
		effective masses and the spin-orbit and crystal-field splitting energies within 		the quasi-cubic approximation.  The effective $A$-band masses 
		derived in the present article ($m_\tn{A}^{\parallel}=1.88\,m_0$ and 
		$m_\tn{A}^{\perp}=0.33\,m_0$) agree very well with the 
		experimental values derived by Rodina \textsl{et al.} 
		($m_\tn{A}^{\parallel}=1.76\,m_0$ and $m_\tn{A}^{\perp}=0.35\,m_0$). 
		Adopting their values for the spin-orbit and crystal-field splitting parameters 		($\Delta_\tn{SO}=0.019$\,eV, $\Delta_\tn{CR}=0.010$\,eV), we 
		also find good agreement for the $B$- and $C$-band masses. 
		(see Tab.~\ref{tab:exp})

                \paragraph*{Wurtzite AlN.} The available experimental data on 
		the band dispersion in wz-AlN is limited to the effective 
		electron mass, which has been determined to be in the range 
		of $0.29$ to $0.45$\,$m_0$.\cite{silveira2004}  The \gw values 
		of $m_\tn{e}^{\parallel}=0.32\,m_0$ and $m_\tn{e}^\perp=0.33\,m_0$ 
		fall within this range.

                \paragraph*{Wurtzite InN.} Experimentally derived effective electron
		masses in wz-InN scatter over a wide range (see Tab.~\ref{tab:exp}). The
		most reliable seem to be those reported by Wu \textsl{et
		al.}\cite{Wu/etal:2002} and Fu \textsl{et al.},\cite{fu2004} since they explicitly account for the high carrier 
		concentration of their samples and the non-parabolicity of the CB in their analysis.  
		Their effective electron masses of $0.05\,m_0$\cite{fu2004} and 
		$0.07\,m_0$\cite{Wu/etal:2002} in conjunction with values 
		for $E_\tn{P}$ of $9.7$\,eV and $10$\,eV, respectively, are in good agreement with those derived from the \gw calculations ($m_\tn{e}^{\parallel}=0.065\,m_0$, $m_\tn{e}^\perp=0.068\,m_0$ and $E_\tn{P}^{\parallel}=8.7$\,eV, $E_\tn{P}^\perp=8.8$\,eV).  Our calculations also predict an anisotropy of the electron masses of about 5\,\%. A similar anisotropy has been reported by Hofmann \textsl{et al.}\cite{hofmann2006} (see Tab.\ref{tab:exp})  

        \subsubsection{\label{sec:other_par} Other parameter sets}
		\paragraph*{Local density approximation (LDA).}\strut\\
		For means of comparison we have also derived band parameters 
		from LDA and OEPx(cLDA) calculations in the same way as for the \gw data.  
		LDA band structures are frequently employed for fitting 
		parameter sets,\cite{kim1997,suzuki1995} but we will 
		demonstrate here, that the LDA is not suitable to consistently determine 
		all parameters for the group-III-nitrides accurately.  
		The parameters derived from the LDA band structures are listed 
		in Tab.~\ref{tab:exp} for the wurtzite phases of GaN and AlN.  
		Since the LDA predicts InN to be metallic, no LDA band 
		parameters could be derived  for InN.

		The effective electron masses of GaN in LDA are smaller than 
		in \gw and the experiment.  The effective electron masses of a 
		given material are, to a first approximation, proportional 
		to the fundamental band gap.\cite{nag2003}  Thus the 
		underestimation of the effective electron masses in LDA is  
		to some degree a
		natural side effect of the underestimation of the fundamental 
		band gap. Additional factors (e.g. self-interaction) 
		contributing to the deviation of 
		the LDA band structure from the quasiparticle one were alluded 
		to in section \ref{sec:OEPx_GW}.

		The $A$-band hole masses in LDA show an increased 
		anisotropy; the deviation from the experimental values 
		increases.

		Despite the fact that the band gap of AlN is also significantly smaller in LDA 
		than in \gw, it is still large, i.e., well above $4$\,eV.  
		Therefore, an effect on the absolute values of the effective electron 
		masses is not visible, but  the LDA predicts an anisotropy of the electron 
		masses, with the opposite sign compared to the \gw calculations.   

		\paragraph*{OEPx(cLDA).}\strut\\
		As alluded to in section \ref{sec:OEPx_GW}, band gaps in the OEPx(cLDA)
		approach open compared to LDA (cf Tab.~\ref{tab:gaps}). Following
		the proportionallity relationship between the direct band gap and the conduction band effective mass, the latter should increase in OEPx(cLDA). This is
		indeed the case, as Tab.~\ref{tab:exp} demonstrates. They are, however,
		also larger than the conduction-band effective masses in the \gw approach,
		despite the fact that only in InN the OEPx(cLDA) band gap is larger than
		that in \gw. We attribute this behaviour to the approximate treatment of 
		correlation in the OEPx(cLDA), which aversely affects the band dispersion
		as explained in Section~\ref{sec:comp_GWs}.
		We thus do not recommend the use of the OEPx or the OEPx(cLDA) 
		approach alone for the determination of band parameters. 
		
                \paragraph*{LDA-plus-correction (LDA+C).}\strut\\
		In the LDA+C approach \cite{christensen1984} delta-function 
		potentials are added at the atomic sites, which artificially 
		push $s$-like wave functions upwards in energy. As a consequence,
		the band gaps open, due to the admixture of cation $s$-states in 
		the conduction band. The potentials have to be fitted to 
		available experimental data, such as the fundamental band gaps 
		and can be applied in an all-electron 
		\cite{christensen1984,carrier2005} but also in a 
		pseudopotential framework.\cite{Segev/etal:2007} 
		Carrier and Wei (CW '05)\cite{carrier2005} determined the 
		effective electron and hole masses of GaN, AlN, and InN, using 
		this method. Their results are also given in Tab.~\ref{tab:exp}.

		Their values for the effective electron masses of all three 
                materials are in good agreement with the \gw 
                values. The deviations are larger for the effective hole masses.
                This is not too surprising because the LDA+C approach 
                predominantly affects $s$-derived bands. Since the upper 
                valence bands around $\Gamma$ are mostly of nitrogen 2$p$ 
		character their description will be closer to the LDA level, whereas 
                the conduction bands feel the additional corrections.

                \paragraph*{Empirical pseudopotential method (EPM).}\strut\\
		A semi-empirical way, often used to calculate band parameters, 
                is the empirical pseudo potential method (EPM).\cite{fritsch2003, fritsch2004, ren1999, dugdale2000}  
                In the EPM the full atomic potentials are replaced by those of
		pseudo atoms, whose adjustable parameters are fitted to a set 
		of input band parameters, typically taken from experiments. The resulting band structures can then be used analogously to fit the parameters of a $\kpe$ Hamiltonian. Since the EPM depends sensitively on the input parameters, appreciable scatter in the reported band parameters is observed. (see Tab. V for a selection)

                \paragraph*{Vurgaftman and Meyer.}\strut\\  For non of the 
		group-III nitrides a complete set of band parameters has so far
		been derived from experimental values alone. Therefore, 
		Vurgaftman and Meyer\cite{Vurgaftman/Meyer:2003} have compiled 
		parameter sets comprising experimental and the most 
		reliable theoretical values in the year 2003.

		For wz-GaN, VM'03 recommend the experimental value of the 
		effective electron masses of 
		$m_\tn{e}^{\parallel}=m_\tn{e}^\perp=0.20\,m_0$ and 
		Luttinger-like parameters derived from EPM calculation by 
		Ren~\textsl{et al.}\cite{ren1999}, which yield effective hole 
		masses in good agreement with experimental and 
		the \gw data (see Tab.~\ref{tab:exp}).  The parameter set 
		yields a band structure that agrees well with the \gw band 
		structure for the CB and the two top VBs 
		(see Fig.~\ref{fig:kp_bs_wz}).  It deviates, however, for 
		the $C$ VB (the third valence band counted from the valence band maximum), 
		where the curvatures in the EPM band structure are 
		too large.
		
		Of all the compounds and phases discussed in this article wz-GaN is the best characterized experimentally. The good agreement between our
		quasiparticle band structures and those based on the parameter set
		recommended by VM'03 proves the quality of our \gw band structures.

                For wz-AlN the effective electron masses recommended by VM'03 
		are the averages over several theoretical values; the 
		recommended VB parameters are theoretical values by
		Kim \emph{et al.}\cite{kim1997} derived from LDA calculations. 
		These parameters yield a band 
		structure, which is in good overall agreement with the \gw band 
		structure (see Fig.~\ref{fig:kp_bs_wz}a).  The anisotropy of 
		the effective electron masses, however, has the opposite sign, similar to our own LDA 
		calculations.   
                The similarity between VM'03 (i.e. LDA) and \gw  in the valence band region
		(after adjusting $\Delta_\tn{CR}$) is due to
		the fact that in AlN valence bands are shifted rigidly compared to the LDA,
		as discussed in Section~\ref{sec:comp_GWs}. In OEPx(cLDA) alone, however, the dispersion 
		changes noticeably (similar to what was observed for InN, cf.\ Fig.~\ref{fig:InN_BS})
		giving rise to appreciably different band parameters (Tab.~\ref{tab:exp}).

		For wz-InN, VM'03 recommend the experimental effective electron masses by 
		Wu \textsl{et al.}\cite{Wu/etal:2002} 
		($m_\tn{e}^{\parallel}=m_\tn{e}^\perp=0.07\,m_0$) and the EPM values from 
		Pugh \textsl{et al.}\cite{pugh1999-1} for the VB.  The pseudo potentials used by 
		Pugh \textsl{et al.} were designed to reproduce their LDA calculations, which had 
		been ``scissor corrected'' to the incorrect band gap of $2.0$\,eV.  
		These parameters are therefore to no avail from today's perspective. 

}

\section{\label{sec:conc}Conclusion}        
We have derived consistent and unbiased band parameters for the wurtzite  and 
zinc-blende phases of GaN, AlN, and InN from accurate \gw band structure calculations.
The band parameters are in very good agreement with the available experimental data, 
proving the reliability of the method. 
We also provide reliable values for
those parameters which have not been determined experimentally, such as, e.g.,
the band parameters of the zinc-blende phases of GaN, AlN, and InN or the
$E_\tn{p}$ and VB parameters of wurtzite phases. These parameters are essential
for understanding the physics of these materials. 
We have derived
complete and consistent parameter sets for the description of the band
structures of the group-III nitrides within $\kpe$-theory.  The $\kpe$-method is
widely used for modeling and simulating (opto-)electronic devices. 
The parameters presented in this work overcome the apparent lack of consistent band 
parameter sets for such simulations.

Finally we remark that the combination of the $\kpe$- with the \GnWn method is not restricted to the 4$\times$4 (8$\times$8) $\kpe$ Hamiltonians discussed in this work.  Since we expect \GnWn to provide the same accuracy for the whole Brillouin zone, the parameters for more complex Hamiltonians can be fitted in the same way.

\begin{acknowledgments}
        We would like to acknowledge fruitful discussions with Axel Hoffmann, 
        Peter Kratzer, Martin Fuchs, Christoph Freysoldt and Chris G. Van de Walle. 
        This work was in part funded by the Volkswagen Stiftung/Germany, 
        the DFG through Sfb 296 and Sfb 787, and the EU's 6th framework program through 
        the NANOQUANTA (NMP4-CT-2004-500198)and SANDiE (NMP4-CT-2004-500101) 
        networks of excellence.
\end{acknowledgments}


\begin{thebibliography}{149}
\bibitem[{GW_()}]{GW_SI}
\bibinfo{note}{Contrary to the OEPx approach the $GW$ method is not free from
  self-interaction. A small amount of self-screening is reintroduced by the
  correlation part of the self-energy, which is, however, significantly smaller
  than the self-interaction error of the LDA \cite{Nelson/Bokes/Rinke/Godby}.}
\end{thebibliography}
\appendix
        \section{$\Kpe$-Hamiltonian \label{sec:kp_ham}}
The $\kpe$-Hamiltonian used in the present work is based on the one introduced in 
Ref.~\onlinecite{kane1982} for zinc-blende crystals and its extension to wurtzite 
crystals structures in Refs.~\onlinecite{winkelnkemper2006}, \onlinecite{chuang1996}, 
 \onlinecite{dugdale2000} and \onlinecite{Bir}.  It will be described in the following.

        Neglecting spin-orbit interaction the 8\,x\,8-Hamilton matrix reduces to 4\,x\,4 and can be decomposed into two separate matrices:
        \begin{equation}
		H=H_1 + H_2 \quad.
	\end{equation}
        The matrix $H_1$ represents the pure 4\,x\,4-$\kpe$ description of the conduction and valence band neglecting all remote band contributions.  For wurtzite crystals it is given by
	\begin{widetext}
        \begin{equation}
		H_1=\left(
			\begin{array}{cccc}
				\tilde{E}_\tn{g}+\Delta_\tn{CR}+\frac{\hbar^2 k^2}{2m_0} & i P_2k_\tn{x} &  i P_2k_\tn{y} &  i P_1k_\tn{z} \\
				-i P_2k_\tn{x} & \Delta_\tn{CR}+\frac{\hbar^2 k^2}{2m_0}   &  0& 0 \\
				-i P_2k_\tn{y} & 0 & \Delta_\tn{CR}+\frac{\hbar^2 k^2}{2m_0}   & 0 \\
				-i P_1k_\tn{z} & 0 & 0 & \frac{\hbar^2 k^2}{2m_0} 
			\end{array}
		\right)\quad.
	\end{equation}
	\end{widetext}
        Here, $m_0$ is the free electron mass. $\tilde{E}_\tn{g}$ is identical to the
	fundamental band gap $E_\tn{g}$ for all materials with a positive crystal-field splitting $\Delta_\tn{CR}$, i.e.\ GaN and InN, and $E_\tn{g}+|\Delta_\tn{CR}|$ for materials with negative $\Delta_\tn{CR}$, i.e.\ AlN. 
	The parameters $P_{1/2}$ are proportional to the absolute value of the CB/VB dipole matrix elements at $\Gamma$.  They are customarily expressed in terms of the Kane parameters $E_\tn{P1/2}$:
        \begin{equation}
		P_{1/2}		=	\sqrt{\frac{\hbar^2}{2m_0}E_\tn{P1/2}}\quad .
        \end{equation}
        For zinc-blende crystals $H_1$ simplifies through $P_{1} = P_{2}$ ($E_\tn{P1}=E_\tn{P2})$ and $\Delta_\tn{CR}=0$.

        The matrix $H_2$ describes the influences of all bands not considered explicitly by the 4\,x\,4-method. For wurtzite crystals it is defined by    
	\begin{widetext}
		\begin{equation}
			H_{2}=\left(
				\begin{array}{cccc}
					A'_2\left(k_\tn{x}^2+k_\tn{y}^2\right)+A'_1k_\tn{z}^2 & B_2k_\tn{y}k_\tn{z} & B_2k_\tn{x}k_\tn{z}  &  B_1k_\tn{x}k_\tn{z} \\
 					B_2k_\tn{y}k_\tn{z} & L'_1k_\tn{x}^2+M_1k_\tn{y}^2+M_2k_\tn{z}^2 &  N'_1k_\tn{x}k_\tn{y} & N'_2k_\tn{x}k_\tn{z}-N'_3k_\tn{x} \\
					B_2k_\tn{x}k_\tn{z} & N'_1k_\tn{y}k_\tn{x} & M_1k_\tn{x}^2+L'_1k_\tn{y}^2+M_2k_\tn{z}^2 & N'_2k_\tn{y}k_\tn{z}-N'_3k_\tn{y} \\
 					B_1k_\tn{x}k_\tn{z} & N'_2k_\tn{z}k_\tn{x}+N'_3k_\tn{x}  & N'_2k_\tn{z}k_\tn{x}+N'_3k_\tn{y}  & M_3\left(k_\tn{x}^2+k_\tn{y}^2\right)+L'_2k_\tn{z}^2
				\end{array}
			\right)\quad.
			\label{eq:g2}
		\end{equation}
	\end{widetext}
        The parameters in $H_2$ are defined in Ref.~\onlinecite{kane1982} in terms of optical matrix elements for zinc-blende crystals. The corresponding definitions for wurtzite crystals can be found in, e.g., Ref.~\onlinecite{chuang1996}.  
        The parameters are related to the more commonly used effective electron masses, $m_\tn{e}^{\parallel}$ and  $m_\tn{e}^{\perp}$, and Luttinger-like parameters, $A_\tn{i}$, by
	\begin{eqnarray}
		A'_1	&	=	&	\frac{\hbar^2}{2}\left(\frac{1}{m_\tn{e}^{\parallel}}-\frac{1}{m_{0}}\right)-\frac{P_1^2}{E_\tn{g}}\quad,\nonumber\\
		A'_2	&	=	&	\frac{\hbar^2}{2}\left(\frac{1}{m_\tn{e}^{\perp}}-\frac{1}{m_{0}}\right)-\frac{P_2^2}{E_\tn{g}}\quad,\nonumber\\
		L'_1	&	=	&	\frac{\hbar^2}{2m_0}\left(A_2+A_4+A_5-1\right)+\frac{P_1^2}{E_\tn{g}}\quad,\nonumber\\
		L'_2	&	=	&	\frac{\hbar^2}{2m_0}\left(A_1-1\right)+\frac{P_2^2}{E_\tn{g}}\quad,\nonumber\\
		M_1	&	=	&	\frac{\hbar^2}{2m_0}\left(A_2+A_4-A_5-1\right)\quad,\nonumber\\
		M_2 	&	=	&	\frac{\hbar^2}{2m_0}\left(A_1+A_3-1\right)\quad,\nonumber\\
		M_3 	&	=	&	\frac{\hbar^2}{2m_0}\left(A_2-1\right)\quad,\nonumber\\
		N_1'	&	=	&	\frac{\hbar^2}{2m_0}2A_5+\frac{P_1^2}{E_\tn{g}}\quad,\nonumber\\
		N_2'	&	=	&	\frac{\hbar^2}{2m_0}\sqrt{2}A_6+\frac{P_1P_2}{E_\tn{g}}\quad,\nonumber\\
		N_3'	&	=	&	i\sqrt{2}A_7\quad.
	\end{eqnarray}
        The corresponding relations for zinc-blende crystals are
	\begin{eqnarray}
        \left(A'_1=A'_2=  \right)A'	&	=	&	\frac{\hbar^2}{2}\left(\frac{1}{m_\tn{e}}-\frac{1}{m_{0}}\right)-\frac{P^2}{E_\tn{g}}\quad,\nonumber\\
	\left(L'_1=L'_2=  \right)L'	&	=	&	-\frac{\hbar^2}{2m_0}\left(\gamma_1+4\gamma_2\right)+\frac{P^2}{E_\tn{g}}\quad,\nonumber\\
	\left(M_1=M_2=M_3=\right)M	&	=	&	-\frac{\hbar^2}{2m_0}\left(\gamma_1-2\gamma_2\right)\quad,\nonumber\\
	\left(N'_1=N'_2=  \right)N'	&	=	&	-\frac{\hbar^2}{2m_0}6\gamma_3+\frac{P^2}{E_\tn{g}}\quad,\nonumber\\
        N'_3            &       =       &       0\quad.
	\end{eqnarray}
                Here, $m_\tn{e}$ denotes the electron effective mass and $\gamma_\tn{i}$ the Luttinger parameters.

                The parameters $B_{1/2}$ occur due to the lack of inversion symmetry in
		zinc-blende and wurtzite crystals.  Their inclusion in the $\kpe$-Hamiltonian does not
		yield a noticeable improvement of the fit results.  Therefore, they
		have been omitted throughout this work. 

\section{Effective hole masses \label{sec:meff} }

The equations, connecting the effective hole masses to the Luttinger(-like) parameters are in detail\\
---for wurtzite crystals---:\cite{chuang1996}
\begin{eqnarray*}
m_0/m_\tn{A}^\parallel	& = & -(A_1+A_3) \quad,\\	
m_0/m_\tn{A}^\perp	& = & -(A_2+A_4) \quad,\\	
m_0/m_\tn{B}^\parallel	& = & -\left[A_1+\left(\frac{E_\tn{B}}{E_\tn{B}-E_\tn{C}}\right)A_3\right] \quad,\\	
m_0/m_\tn{B}^\perp	& = & -\left[A_2+\left(\frac{E_\tn{B}}{E_\tn{B}-E_\tn{C}}\right)A_4\right] \quad,\\	
m_0/m_\tn{C}^\parallel	& = & -\left[A_1+\left(\frac{E_\tn{C}}{E_\tn{C}-E_\tn{B}}\right)A_3\right] \quad,\\	
m_0/m_\tn{C}^\perp	& = & -\left[A_2+\left(\frac{E_\tn{C}}{E_\tn{C}-E_\tn{B}}\right)A_4\right]\quad, 	
\end{eqnarray*}
with
\begin{eqnarray*}
E_\tn{B}	& = & \frac{\Delta_\tn{CR}-\Delta_\tn{SO}/3}{2}\\
		&   & +\sqrt{\left(\frac{ \Delta_\tn{CR}-\Delta_\tn{SO}/3}{2}\right)^2+2\left(\frac{\Delta_\tn{SO}/3}{2}\right)^2} \quad, \\
E_\tn{C}	& = & \frac{\Delta_\tn{CR}-\Delta_\tn{SO}/3}{2}\\
		&   & -\sqrt{\left(\frac{ \Delta_\tn{CR}-\Delta_\tn{SO}/3}{2}\right)^2+2\left(\frac{\Delta_\tn{SO}/3}{2}\right)^2} \quad. 
\end{eqnarray*}
For AlN the indices  A, B, and C have to be interchanged: $\tn{A}\rightarrow\tn{B}$, $\tn{B}\rightarrow\tn{C}$, $\tn{C}\rightarrow\tn{A}$.\\
For zinc blende crystals it follows:\cite{vurgaftman2001}
\begin{eqnarray*}
m_0/m_\tn{hh}^\tn{[001]}	& = & \gamma_1-2\gamma_2 \quad,\\	
m_0/m_\tn{hh}^\tn{[110]}	& = & \frac{1}{2}(2\gamma_1-\gamma_2-3\gamma_3) \quad,\\
m_0/m_\tn{hh}^\tn{[111]}	& = & \gamma_1-2\gamma_3 \quad,\\	
m_0/m_\tn{lh}^\tn{[001]}	& = & \gamma_1+2\gamma_2 \quad,\\	
m_0/m_\tn{lh}^\tn{[110]}	& = & \frac{1}{2}(2\gamma_1+\gamma_2+3\gamma_3) \quad,\\
m_0/m_\tn{lh}^\tn{[111]}	& = & \gamma_1+2\gamma_3 \quad,\\	
m_0/m_\tn{so}			& = & \gamma_1-\frac{E_\tn{P}\Delta_\tn{SO}}{3E_\tn{g}(E_\tn{g}+\Delta_\tn{SO})} \quad.\\	
\end{eqnarray*}

\end{document}